\documentclass[10pt,onecolumn,aps,prd,preprintnumbers,showpacs,superscriptaddress,nofootinbib,amsmath,amssymb,floats,floatfix,showkeys,notitlepage,longbibliography]{revtex4-2}

\usepackage{orcidlink}
\usepackage{comment}
\usepackage{lipsum}
\usepackage{graphicx}
\usepackage{subfigure}
\usepackage{palatino}
\usepackage{sans}
\usepackage{hyperref}
\hypersetup{colorlinks=true,linkcolor=blue,urlcolor=blue,citecolor=blue}
\usepackage[toc,page]{appendix}
\usepackage[normalem]{ulem}
\usepackage{adjustbox}
\usepackage{latexsym}
\usepackage{amsmath}
\usepackage{amssymb}
\usepackage{amsfonts}
\usepackage{dcolumn}
\usepackage{bm}
\usepackage{tikz}
\usetikzlibrary{decorations.pathmorphing}
\usepackage{bigints}
\usepackage{array,tabularx,multirow,booktabs}
\usepackage[tracking=true]{microtype}
\usepackage{soul} %for highlighting
\SetTracking{}{500}
\SetTracking{encoding={*}, shape=sc}{40}
\UseRawInputEncoding %for inputenc error%
\allowdisplaybreaks

\usepackage[utf8]{inputenc}
\usepackage{algorithm}
\usepackage{algorithmicx}
\usepackage{algpseudocode}
\usepackage{xcolor} % For colored text if needed

\begin{document} \sloppy

\title{Gravitational Black Hole Shadow Spectroscopy}

\author{Reggie C. Pantig \orcidlink{0000-0002-3101-8591}} 
\email{rcpantig@mapua.edu.ph}
\affiliation{Physics Department, School of Foundational Studies and Education, Map\'ua University, 658 Muralla St., Intramuros, Manila 1002, Philippines.}

\author{Ali \"Ovg\"un \orcidlink{0000-0002-9889-342X}}
\email{ali.ovgun@emu.edu.tr}
\affiliation{Physics Department, Faculty of Arts and Sciences, Eastern Mediterranean University, Famagusta, 99628 North
Cyprus via Mersin 10, Turkiye.}

\begin{abstract}
In this work, we develop a generalized perturbative framework for gravitational shadows in static, spherically symmetric spacetimes. Building upon the recent two-parameter perturbative framework of Kobialko et al. \cite{Kobialko:2024zhc}, this work extends the expansion in particle energy and metric deviation to encompass arbitrary, simultaneous deformations of all metric functions. By relaxing the common restriction of a fixed area radius $(\beta(r) = r^2)$, our formalism applies to a significantly broader class of alternative gravity theories and exotic compact objects. We derive analytical formulae for the massive shadow radius up to the second order in the deformation parameter, explicitly revealing the phenomenological signatures that arise from the coupling between temporal and spatial metric perturbations. The key result is that the distinct energy dependence of the massive shadow provides a powerful method to disentangle these different types of geometric deformations, breaking observational degeneracies inherent in the photon shadow alone. We demonstrate this principle with applications to traversable wormholes and canonical scalar-tensor solutions, showing how each produces a unique, distinguishable energy-dependent fingerprint. This generalized framework provides a robust, theory-agnostic tool for testing strong-field gravity. It offers a clear methodology for reconstructing metric parameters from potential multi-messenger observations of massive particle shadows.
\end{abstract}


\maketitle

\section{Introduction}\label{intro}
The direct imaging of horizon-scale structure in M87* and Sgr A* by the Event Horizon Telescope (EHT) has transformed the black-hole shadow from a theoretical curiosity into a precision observable for strong-field gravity \cite{EventHorizonTelescope:2019dse,EventHorizonTelescope:2022wkp}. Together with decades of theoretical groundwork on shadow formation and ray-tracing \cite{Synge:1966okc,Cunningham_1972,Bardeen:1973tla,Luminet:1979nyg,Chandrasekhar:1985kt,Falcke:1999pj,Bambi:2015kza,Psaltis:2018xkc,Perlick:2021aok,Vagnozzi:2022moj}, these results have catalyzed a broad program to test General Relativity (GR) and its alternatives at horizon scales, while also highlighting systematic challenges and degeneracies in interpreting a single shadow diameter as a stand-in for spacetime geometry \cite{Vagnozzi:2020quf}. Looking ahead, the next generation aims to resolve substructures, such as narrow photon rings, and measure their separations and temporal evolution, thereby unlocking new insights into the metric beyond the average ring size \cite{Ayzenberg:2023hfw,Tiede:2022grp,Aratore:2024bro,Perlick:2015vta}.

At the geometric level, shadow formation is governed by photon surfaces/light rings and, more generally, by families of unstable circular orbits \cite{Claudel:2000yi,Decanini:2010fz,Shoom:2017ril,Hod:2020pim,Hod:2023jmx}. In stationary spacetimes, these include spherical photon orbits in Kerr \cite{Teo:2003ltt,Hod:2012ax,Hod:2013mgr,Tsukamoto:2014tja,Tsukamoto:2017fxq}. A key lesson from recent analyses is that multiple, physically distinct geometries can cast nearly identical shadows (shadow degeneracy) unless supplementary information or more differential observables are used \cite{Cunha:2018acu,Cunha:2018gql,Lima:2021las,Abdujabbarov:2015xqa}. These studies motivated parametrized descriptions of deviations from Schwarzschild/Kerr to link shadow features to metric coefficients in a model-agnostic way \cite{Konoplya:2016jvv,Cardoso:2014rha,Johannsen:2011dh,Younsi:2016azx,Bronnikov:2021liv,Konoplya:2021slg}. Environmental and dynamical effects like plasma dispersion, accretion variability, cosmological backgrounds can also bias or reshape the silhouette \cite{Perlick:2015vta,Kobialko:2025sls,Koga:2022dsu,Solanki:2022glc,Tsupko:2019mfo,Tsukamoto:2014tja,Wang:2017qhh,Johannsen:2013vgc,Feleppa:2024vdk,Feleppa:2025ejh}.

A complementary direction is to go beyond massless probes. Massive test particles (and effectively massive photons in dispersive media) admit energy-dependent unstable circular orbits whose projection defines an energy-dependent massive shadow. Strong-deflection and ring-structure analyses for timelike geodesics have recently been developed and sharpen this perspective \cite{Feleppa:2024kio,Qiao:2022hfv,Paithankar:2023ofw,Lu:2019zxb}. Building on this idea, Vertogradov et al. \cite{Vertogradov:2024qpf,Vertogradov:2024dpa,Vertogradov:2024jzj} introduced general analytic approaches that connect deformations of the metric to photon sphere and shadow radii, including mass-dependent variations. Similarly, analytical formalisms have been developed to connect metric deformations to a wider range of observables, from weak-field particle deflection to strong-field shadow sizes, for systematic tests with EHT data \cite{Pantig:2024kqy}. In parallel, Kobialko et al. \cite{Kobialko:2024zhc} improved Vertogradov's et al. work \cite{Vertogradov:2024dpa} and  formulated a perturbation theory for gravitational shadows in static, spherically symmetric spacetimes, with recent extensions to plasma-immersed disks and emission spectra \cite{Kobialko:2025sls}. These developments reinforce the view that spectral information, which is how a shadow changes with probe energy/frequency, can break degeneracies intrinsic to a single, achromatic silhouette.

In this work, we develop a generalized, theory-agnostic perturbative framework for gravitational shadow spectroscopy in static, spherically symmetric spacetimes. Our approach extends dual expansions in (i) the particle mass-to-energy ratio and (ii) metric deviations by allowing simultaneous deformations of all metric functions, notably relaxing the usual area-radius gauge $\beta(r)=r^{2}$. In contrast to many treatments where only $g_{tt}$ is perturbed, our formalism keeps coupled perturbations of both temporal and spatial sectors and shows explicitly how their interplay imprints a distinctive, energy-dependent signature on the massive shadow. In particular, we derive closed expressions for the first- and second-order corrections to the shadow radius and exhibit coupling terms that are invisible to photon-only analyses. The outcome is a practical recipe to disentangle temporal vs. spatial deformations by scanning the shadow with particles (or effectively massive photons in plasma), thereby mitigating shadow degeneracy \cite{Perlick:2015vta,Perlick:2021aok,Cunha:2018acu,Lima:2021las}.

We illustrate the method on (i) a Simpson–Visser black-bounce wormhole \cite{Simpson:2018tsi} and (ii) the Fisher–Janis–Newman–Winicour scalar solution from scalar-tensor theory \cite{Sau:2020xau}, which naturally modifies both $g_{tt}$ and the area function and thus lies outside the scope of fixed-$\beta$ perturbations. The framework also interfaces with broader probes such as lensing observables, QNMs in the eikonal limit, and neutrino/particle propagation in modified gravity and matter-supported spacetimes \cite{Cunha:2017qtt,Churilova:2019jqx,Hennigar:2018hza,Konoplya:2020jgt,Vagnozzi:2022moj,Allahyari:2019jqz,Kuang:2022xjp,Okyay:2021nnh,Pantig:2022gih,Pantig:2022sjb,Pantig:2022qak,Ovgun:2024zmt,Karshiboev:2024xxx,Alloqulov:2024olb,Abdujabbarov:2016hnw,Mustafa:2022xod,Rayimbaev:2022hca,Yang:2023tip,Adler:2022qtb,Virbhadra:1999nm,Virbhadra:2007kw,Virbhadra:2022iiy,Heidari:2024bkm}. In this sense, shadow spectroscopy complements ongoing efforts with the EHT to resolve photon-ring substructure and to combine electromagnetic and gravitational-wave information for horizon-scale tests \cite{Ayzenberg:2023hfw,Tiede:2022grp}.

Section \ref{sec2} sets up the generalized metric expansion and derives energy-dependent conditions for the massive-particle sphere (MPS) and the shadow radius. Section \ref{sec3} performs the perturbative expansion to second order, isolating linear and nonlinear couplings of temporal and spatial perturbations. Section \ref{sec4} presents the worked examples and highlights distinctive, energy-dependent fingerprints. We conclude in Section \ref{conc} with observational prospects, connections to time-dependent and dispersive environments \cite{Koga:2022dsu,Solanki:2022glc,Perlick:2015vta,Kobialko:2025sls}, and propose some research directions. Throughout this paper, we used geometrized units by setting $G=c=1$, and the metric signature $(-,+,+,+)$.

\section{Generalized Formalism for Massive Particle Orbits and Shadows} \label{sec2}
To construct a perturbation theory capable of describing a broad class of deviations from the Schwarzschild geometry, we begin by relaxing the gauge choice imposed in the examples of \cite{Kobialko:2024zhc}. We consider the most general static, spherically symmetric four-dimensional spacetime, whose geometry is described by the line element \cite{Perlick:2021aok,Vagnozzi:2022moj}
\begin{equation} \label{eq_gen_metric}
    ds^2 = -\alpha(r, \delta) dt^2 + \gamma(r, \delta) dr^2 + \beta(r, \delta) \left(d\theta^2 + \sin^2\theta d\phi^2\right),
\end{equation}
where $\alpha$, $\beta$, and $\gamma$ are arbitrary functions of the radial coordinate $r$. The parameter $\delta$ is a dimensionless quantity that controls the deviation from a chosen background spacetime, which we take to be the Schwarzschild solution. For $\delta=0$, we recover the familiar Schwarzschild metric.

\subsection{The Generalized Metric Ansatz} \label{ssec2.1}
Our central methodological step is to introduce a fully generalized perturbative expansion for all three metric functions. We express each function as a power series in the small deformation parameter $\delta$, expanding around the Schwarzschild background:
\begin{align}
    \alpha(r, \delta) &= \alpha_0(r) + \delta \alpha_1(r) + \delta^2 \alpha_2(r) + \mathcal{O}(\delta^3), \label{eq_alpha_exp} \\
    \beta(r, \delta) &= \beta_0(r) + \delta \beta_1(r) + \delta^2 \beta_2(r) + \mathcal{O}(\delta^3), \label{eq_beta_exp} \\
    \gamma(r, \delta) &= \gamma_0(r) + \delta \gamma_1(r) + \delta^2 \gamma_2(r) + \mathcal{O}(\delta^3). \label{eq_gamma_exp}
\end{align}

Kobialko et al. \cite{Kobialko:2024zhc} develop a shadow perturbation theory on a Schwarzschild background while fixing the areal gauge $\beta=r^{2}$, so that $\beta_{1}=\beta_{2}=0$ and the analysis effectively tracks deformations of the temporal sector $\alpha(r)$ (with examples such as Reissner-Nordstr\"om treated under this restriction). Our contribution is to remove this assumption. We introduce a generalized metric ansatz that perturbs all metric functions, given by Eq. \eqref{eq_alpha_exp}-\eqref{eq_gamma_exp}, thereby allowing departures from the Schwarzschild area radius. This captures spacetimes where the area of spheres is intrinsically modified (e.g., wormholes, scalar–tensor solutions, regular black holes), and it makes the massive-particle sphere (MPS) and shadow explicitly sensitive to couplings between temporal and spatial perturbations. Consequently, already at first-order the shadow depends on a distinct linear combination of $\{\alpha_{1},\beta_{1}\}$, while at second order non-linear $\alpha$–$\beta$ couplings appear, indicating signatures that are absent in the fixed-$\beta$ framework of \cite{Kobialko:2024zhc}.

The zeroth-order terms correspond to the standard Schwarzschild metric components in Schwarzschild coordinates, given by
\begin{equation}
    \alpha_0(r) = 1 - \frac{2M}{r}, \quad \beta_0(r) = r^2, \quad \gamma_0(r) = \left(1 - \frac{2M}{r}\right)^{-1},
\end{equation}
where $M$ is the ADM mass of the background spacetime.

The functions $\alpha_i(r)$, $\beta_i(r)$, and $\gamma_i(r)$ for $i \ge 1$ represent arbitrary, theory-agnostic perturbation profiles. This ansatz is significantly more general than the framework used for the examples in \cite{Kobialko:2024zhc}, which was largely restricted to the case where $\beta_1(r) = \beta_2(r) = 0$. By allowing for non-zero perturbations in $\beta(r, \delta)$, we can now systematically investigate spacetime geometries where the area of spheres, $A(r) = 4\pi\beta(r)$, deviates from the simple $4\pi r^2$ relation characteristic of the Schwarzschild gauge. This generalization is crucial for analyzing solutions from alternative theories of gravity where such modifications naturally arise.

As established in \cite{Kobialko:2024zhc} and will be reaffirmed in the next section, the equations governing the shadow radius for an asymptotic observer depend exclusively on the functions $\alpha(r)$ and $\beta(r)$. The function $\gamma(r)$ does not enter into the final expression for the shadow boundary. Nevertheless, we retain its expansion in our ansatz for completeness and to facilitate potential future extensions of this work, such as the analysis of the stability of massive particle orbits or the calculation of perihelion precession, where the full metric structure is required.

\subsection{Generalized Conditions for Massive Particle Spheres (MPS)} \label{ssec2.2}
The gravitational shadow is delineated by light rays or particle trajectories that asymptotically approach unstable circular orbits. These orbits foliate a timelike hypersurface known as the MPS. To determine the location of the MPS in our generalized spacetime, we analyze the geodesic motion of a particle with mass $m$.

Let us begin with the Lagrangian for a test particle moving in the spacetime described by Eq. \eqref{eq_gen_metric}. This is expressed as
\begin{equation}
    \mathcal{L} = \frac{1}{2} g_{\mu\nu} \dot{x}^\mu \dot{x}^\nu = \frac{1}{2} \left( -\alpha(r, \delta) \dot{t}^2 + \gamma(r, \delta) \dot{r}^2 + \beta(r, \delta) (\dot{\theta}^2 + \sin^2\theta \dot{\phi}^2) \right),
\end{equation}
where the overdot denotes differentiation with respect to an affine parameter. The staticity and spherical symmetry of the metric give rise to two Killing vectors, $\xi^\mu_{(t)} = (1, 0, 0, 0)$ and $\xi^\mu_{(\phi)} = (0, 0, 0, 1)$. Along the geodesic, these symmetries imply the conservation of the particle's energy $E$ and angular momentum $L_z$ via
\begin{align}
    E &= -p_t = -g_{t\mu} \dot{x}^\mu = \alpha(r, \delta) \dot{t}, \\
    L_z &= p_\phi = g_{\phi\mu} \dot{x}^\mu = \beta(r, \delta) \sin^2\theta \dot{\phi}.
\end{align}
Without loss of generality, we can restrict our analysis to the equatorial plane ($\theta = \pi/2$, $\dot{\theta}=0$), where the total angular momentum is $L = L_z$. The normalization condition for the four-velocity of a massive particle, $g_{\mu\nu} \dot{x}^\mu \dot{x}^\nu = -m^2$, can now be used to derive the radial equation of motion. Substituting the conserved quantities, we find
\begin{equation}
    -\frac{E^2}{\alpha(r, \delta)} + \gamma(r, \delta) \dot{r}^2 + \frac{L^2}{\beta(r, \delta)} = -m^2.
\end{equation}
Rearranging this expression to isolate the radial kinetic term yields
\begin{equation}
    \gamma(r, \delta) \dot{r}^2 = \frac{E^2}{\alpha(r, \delta)} - \frac{L^2}{\beta(r, \delta)} - m^2.
\end{equation}
Following the convention of \cite{Kobialko:2024zhc}, we can define an effective potential $V(r, \delta)$ that governs the radial motion as
\begin{equation} \label{eq_V_eff}
    E^{-2} \gamma(r, \delta) \dot{r}^2 = V(r, \delta) \equiv \frac{1}{\alpha(r, \delta)} - \frac{l}{\beta(r, \delta)} - \epsilon,
\end{equation}
where we have introduced the conserved specific energy squared $\epsilon = m^2/E^2$ and the specific angular momentum squared $l = L^2/E^2$.

The MPS are, by definition, located at radii $r$ where stable or unstable circular orbits are possible. A circular orbit requires the radial velocity and radial acceleration to vanish simultaneously. In the effective potential formalism, this corresponds to the conditions that the particle rests at an extremum of the potential:
\begin{equation}
    V(r, \delta) = 0 \quad \text{and} \quad V_{,r}(r, \delta) \equiv \frac{\partial V}{\partial r} = 0.
\end{equation}
Applying these two conditions to the effective potential in Eq. \eqref{eq_V_eff} gives us the following system of equations:
\begin{align}
    \frac{1}{\alpha} - \frac{l}{\beta} - \epsilon &= 0, \label{eq_V_zero} \\
    -\frac{\alpha_{,r}}{\alpha^2} + \frac{l \beta_{,r}}{\beta^2} &= 0. \label{eq_V_prime_zero}
\end{align}
Here, the subscript '$,r$' denotes a partial derivative with respect to $r$, and we have suppressed the explicit dependence on $(r, \delta)$ for notational clarity.

This system of two algebraic equations can be solved for the orbit parameters $\epsilon$ and $l$. From Eq. \eqref{eq_V_prime_zero}, we first solve for the specific angular momentum squared, $l$:
\begin{equation}
    l = \frac{\beta^2}{\alpha^2} \frac{\alpha_{,r}}{\beta_{,r}}.
\end{equation}
Substituting this expression back into Eq. \eqref{eq_V_zero} allows us to solve for the specific energy squared, $\epsilon$. We find
\begin{equation}
    \epsilon = \frac{1}{\alpha} - \frac{1}{\beta} \left( \frac{\beta^2}{\alpha^2} \frac{\alpha_{,r}}{\beta_{,r}} \right) = \frac{1}{\alpha} \left( 1 - \frac{\beta}{\alpha} \frac{\alpha_{,r}}{\beta_{,r}} \right).
\end{equation}

Thus, the fundamental equations defining the radius $r$ of the MPS remain formally identical to those derived in \cite{Kobialko:2024zhc}, and these are
\begin{align}
    \epsilon &= \frac{1}{\alpha(r, \delta)} \left( 1 - \frac{\beta(r, \delta)}{\alpha(r, \delta)}  \frac{\alpha_{,r}(r, \delta)}{\beta_{,r}(r, \delta)} \right), \label{eq_epsilon_gen} \\
    l &= \frac{\beta(r, \delta)^2}{\alpha(r, \delta)^2}  \frac{\alpha_{,r}(r, \delta)}{\beta_{,r}(r, \delta)}. \label{eq_l_gen}
\end{align}
The crucial difference, however, lies in the interpretation of these equations. With our generalized metric ansatz from Eqs. \eqref{eq_alpha_exp}-\eqref{eq_gamma_exp}, the functions $\alpha$, $\beta$, and their derivatives are no longer simple functions of $r$ but are perturbative series in $\delta$ with arbitrary coefficient functions $\alpha_i(r)$ and $\beta_i(r)$.

Consequently, Eq. \eqref{eq_epsilon_gen} is no longer a simple algebraic equation for $r$. It is now a highly complex, transcendental equation that implicitly defines the MPS radius as a function of the particle's energy parameter and the metric deformation, $r = r(\epsilon, \delta)$. Critically, this implicit function is now sensitive to perturbations in both the temporal component $\alpha(r, \delta)$ and the angular component $\beta(r, \delta)$ of the metric. This coupling between the perturbations, mediated through the ratio of derivatives $\alpha_{,r}/\beta_{,r}$, is the key feature of our generalized framework and the source of new phenomenology that we will explore in the subsequent sections. The primary task of our perturbative analysis will be to explicitly solve for the function $r(\epsilon, \delta)$ as a power series in $\delta$ in order to ultimately determine the corresponding corrections to the gravitational shadow.

\subsection{Generalized shadow radius} \label{ssec2.3}
Having established the conditions that determine the location of the MPS, we now turn to the primary observable: the gravitational shadow. The shadow boundary is formed by the trajectories of particles that originate on the MPS and reach the observer. To calculate its size, we must relate the properties of the particle on its circular orbit (specifically, its specific angular momentum $l$ and energy parameter $\epsilon$) to the apparent angle on a distant observer's celestial sphere.

Let us consider a static observer located at a large but finite coordinate radius $\bar{r}$ in the equatorial plane ($\theta = \pi/2$). In their local frame, this observer measures the components of the incoming particle's four-velocity using an orthonormal tetrad $\{e^\mu_{(a)}\}$ adapted to the static, spherically symmetric coordinates. For the metric given by Eq. \eqref{eq_gen_metric}, a suitable tetrad is
\begin{align}
e^\mu_{(t)} &= \frac{1}{\sqrt{\alpha(\bar{r}, \delta)}} (1, 0, 0, 0), \nonumber \\
e^\mu_{(r)} &= \frac{1}{\sqrt{\gamma(\bar{r}, \delta)}} (0, 1, 0, 0), \nonumber \\
e^\mu_{(\phi)} &= \frac{1}{\sqrt{\beta(\bar{r}, \delta)}} (0, 0, 0, 1).
\end{align}
The components of the particle's four-velocity $\dot{x}^\mu$ in this local Lorentz frame are given by $u^{(a)} = e^\mu_{(a)} g_{\mu\nu} \dot{x}^\nu$. Using the conserved energy $E = \alpha \dot{t}$ and angular momentum $L = \beta \dot{\phi}$, the locally measured temporal and azimuthal components of the four-velocity are given by
\begin{align}
    u^{(t)} &= e^\mu_{(t)} p_t = -\frac{E}{\sqrt{\alpha(\bar{r}, \delta)}}, \\
    u^{(\phi)} &= e^\mu_{(\phi)} p_\phi = \frac{L}{\sqrt{\beta(\bar{r}, \delta)}}.
\end{align}
The normalization condition in the local frame is $\eta_{ab} u^{(a)} u^{(b)} = -(u^{(t)})^2 + (u^{(r)})^2 + (u^{(\phi)})^2 = -m^2$. Substituting the components above, we find
$$-\frac{E^2}{\alpha(\bar{r}, \delta)} + (u^{(r)})^2 + \frac{L^2}{\beta(\bar{r}, \delta)} = -m^2.$$
The apparent angle $\Theta$ of an incoming particle trajectory on the observer's sky, relative to the radial direction, is defined by the ratio of the tangential and radial momentum components. Following the procedure in \cite{Kobialko:2024zhc}, we can relate the angle to the conserved quantities:
\begin{equation}
    \sin^2 \Theta = \frac{(u^{(\phi)})^2}{(u^{(r)})^2 + (u^{(\phi)})^2} = \frac{L^2/\beta(\bar{r}, \delta)}{E^2/\alpha(\bar{r}, \delta) - m^2}.
\end{equation}
Dividing the numerator and denominator by $E^2$ and using the definitions $\epsilon = m^2/E^2$ and $l = L^2/E^2$, we arrive at the angular size of the shadow as seen by an observer at some finite radius $\bar{r}$:
\begin{equation} \label{eq_sin_theta_finite}
    \sin^2 \Theta = \frac{l / \beta(\bar{r}, \delta)}{1/\alpha(\bar{r}, \delta) - \epsilon} = \frac{\alpha(\bar{r}, \delta)}{\beta(\bar{r}, \delta)} \frac{l}{1 - \alpha(\bar{r}, \delta) \epsilon}.
\end{equation}
Such an expression connects the observed angle $\Theta$ to the specific angular momentum $l$ of the particle tracing the shadow's edge. Since this particle must have originated from an MPS orbit, we substitute the expression for $l$ from Eq. \eqref{eq_l_gen}, which is evaluated at the MPS radius $r = r(\epsilon, \delta)$. The result is
\begin{equation} \label{eq_sin_theta_full}
\sin^2\Theta = \frac{\alpha(\bar{r}, \delta)}{\beta(\bar{r}, \delta)}  \frac{1}{1 - \alpha(\bar{r}, \delta) \epsilon}  \left[ \frac{\beta(r, \delta)^2}{\alpha(r, \delta)^2} \frac{\alpha_{,r}(r, \delta)}{\beta_{,r}(r, \delta)} \right]_{r=r(\epsilon,\delta)}.
\end{equation}
We can simplify this expression significantly by using the condition for the MPS radius itself, Eq. \eqref{eq_epsilon_gen}, which relates the orbit parameters. From Eq. \eqref{eq_epsilon_gen}, we have the identity $\frac{\beta}{\alpha} \frac{\alpha_{,r}}{\beta_{,r}} = 1 - \alpha\epsilon$. Substituting this into the expression for $l$ gives $l = \frac{\beta}{\alpha}(1-\alpha\epsilon)$. When this is inserted into Eq. \eqref{eq_sin_theta_finite}, we obtain a more direct relation expressed as
$$ \sin^2\Theta = \frac{\alpha(\bar{r}, \delta)}{\beta(\bar{r}, \delta)}  \frac{\beta(r, \delta)/\alpha(r, \delta)  (1-\alpha(r,\delta)\epsilon)}{1 - \alpha(\bar{r}, \delta) \epsilon}.$$

For most astrophysical applications, the observer is located at a cosmological distance from the compact object, so we are interested in the asymptotic limit $\bar{r} \to \infty$. For an asymptotically flat spacetime, the metric functions at the observer's location approach their Minkowski values:
\begin{equation}
    \lim_{\bar{r}\to\infty} \alpha(\bar{r}, \delta) = 1, \quad \text{and} \quad \lim_{\bar{r}\to\infty} \frac{\beta(\bar{r}, \delta)}{\bar{r}^2} = 1.
\end{equation}
In this limit, the apparent angular size $\Theta$ tends to zero. We define the finite, observable shadow radius $R$ by scaling this small angle with the distance $\bar{r}$, we get
\begin{equation}
    R^2 \equiv \lim_{\bar{r}\to\infty} \left( \bar{r}^2 \sin^2\Theta \right).
\end{equation}
Applying this limit to our expression for $\sin^2\Theta$, we find:
\begin{equation}
    R^2 = \lim_{\bar{r}\to\infty} \left( \bar{r}^2 \frac{1}{\bar{r}^2}  \frac{\beta(r, \delta)}{\alpha(r, \delta)} \frac{1-\alpha(r,\delta)\epsilon}{1 - \epsilon} \right).
\end{equation}
The above yields the final, generalized expression for the squared radius of the massive shadow as seen by an asymptotic observer:
\begin{equation}
    R^2(\epsilon, \delta) = \frac{\beta(r, \delta)}{\alpha(r, \delta)} \frac{1 - \alpha(r, \delta) \epsilon}{1 - \epsilon}, \label{eq_R2_gen}
\end{equation}
where the radius $r$ at which the metric functions are evaluated is itself implicitly defined by the MPS condition, Eq. \eqref{eq_epsilon_gen}.

It is essential to recognize the profound implication of this result. The final formula, Eq. \eqref{eq_R2_gen}, is formally identical to the one derived in \cite{Kobialko:2024zhc}. However, its functional dependence on the underlying physical parameters is now significantly more complex and contains a wealth of new information. In the previous, restricted framework, the dependence on the metric deformation $\delta$ entered primarily through $\alpha(r, \delta)$ and the shift in the MPS radius $r(\epsilon, \delta)$ induced by it. In our generalized framework, the shadow radius $R^2$ now depends on the full set of perturbation functions, $\{\alpha_i(r), \beta_i(r)\}$. The prefactor $\beta/\alpha$ is directly sensitive to perturbations in both the temporal and spatial geometry, while the implicitly defined MPS radius $r(\epsilon, \delta)$ now depends on a complex interplay between the perturbations $\alpha_i$ and $\beta_i$ and their derivatives, as seen in Eq. \eqref{eq_epsilon_gen}. This intricate, coupled dependence is precisely what allows for the possibility of disentangling different types of metric deformations by observing the shadow's size across a spectrum of particle energies $\epsilon$. The primary goal of the following section is to systematically expand this compact expression in the small parameter $\delta$ to extract these phenomenological signatures.

\section{Perturbative Expansion of the Shadow Radius} \label{sec3}
The core of our analysis is to solve for the shadow radius $R^2(\epsilon, \delta)$ as a power series in the small deformation parameter $\delta$. The background for this expansion is the exact solution for the massive shadow in the unperturbed Schwarzschild spacetime, which we denote by $R^2_{\rm MSch}(\epsilon)$. The corresponding radius of the MPS in this background, which we denote by $r_0(\epsilon)$, is also known analytically and was presented in \cite{Kobialko:2024zhc}. The primary challenge lies in systematically computing the corrections to these quantities arising from the generalized metric perturbations introduced in Sec. \ref{ssec2.1}.

\subsection{First-Order Correction to the MPS Radius} \label{ssec3.1}
We begin by calculating the correction to the shadow radius at the first-order in $\delta$. This involves a two-step process: first, we must determine the first-order shift in the MPS radius itself, and second, we use this to find the resulting first-order change in the shadow radius. As we will demonstrate, a remarkable simplification occurs in the second step, rendering the final expression for the shadow correction independent of the explicit form of the MPS radius correction.

The MPS radius $r(\epsilon, \delta)$ is defined implicitly by Eq. \eqref{eq_epsilon_gen}, which we can write schematically as $\epsilon = \mathcal{F}(r(\epsilon, \delta), \delta)$, where
$$\mathcal{F}(r, \delta) = \frac{1}{\alpha(r, \delta)} \left( 1 - \frac{\beta(r, \delta)}{\alpha(r, \delta)} \frac{\alpha_{,r}(r, \delta)}{\beta_{,r}(r, \delta)} \right).$$
To find the first-order correction, we expand the MPS radius as a series in $\delta$ around the background solution $r_0(\epsilon)$ via
\begin{equation}
r(\epsilon, \delta) = r_0(\epsilon) + \delta \, r_1(\epsilon) + \mathcal{O}(\delta^2).
\end{equation}
We now expand the implicit equation $\epsilon = \mathcal{F}(r, \delta)$ to first-order in $\delta$. The left-hand side, $\epsilon$, is independent of $\delta$. The right-hand side is expanded around the point $(r_0, 0)$:
\begin{equation}
\mathcal{F}(r(\epsilon, \delta), \delta) \approx \mathcal{F}(r_0, 0) + \delta \left[ \frac{\partial \mathcal{F}}{\partial r} \bigg|_{(r_0,0)} r_1(\epsilon) + \frac{\partial \mathcal{F}}{\partial \delta} \bigg|_{(r_0,0)} \right] + \mathcal{O}(\delta^2).
\end{equation}
By definition, the zeroth-order term corresponds to the background solution, so $\mathcal{F}(r_0, 0) = \epsilon$. For the entire equation to hold, the term of order $\delta$ must vanish. This gives us a linear equation for the first-order correction to the radius, $r_1(\epsilon)$:
\begin{equation} \label{eq_r1_implicit}
\frac{\partial \mathcal{F}}{\partial r} \bigg|_{(r_0,0)} r_1(\epsilon) + \frac{\partial \mathcal{F}}{\partial \delta} \bigg|_{(r_0,0)} = 0.
\end{equation}
Solving for $r_1(\epsilon)$ yields
\begin{equation} \label{eq_r1_solution}
r_1(\epsilon) = - \frac{ \partial \mathcal{F}/\partial \delta }{ \partial \mathcal{F}/\partial r } \bigg|_{(r_0,0)}.
\end{equation}
The denominator, $\partial \mathcal{F}/\partial r \big|_{(r_0,0)}$, is related to the stability of the background circular orbits. It is proportional to the second derivative of the effective potential, $V_{,rr}$, evaluated for the Schwarzschild background, and is generically non-zero for the unstable orbits that form the shadow. The numerator involves the partial derivatives of the metric perturbation functions, $\alpha_1(r)$ and $\beta_1(r)$, evaluated at $r_0(\epsilon)$. While an explicit (and rather lengthy) expression for $r_1(\epsilon)$ can be derived by computing these derivatives, we will find that it is not required for the first-order correction to the shadow radius itself.

We now compute the first-order correction to the squared shadow radius, $R^2$. The full expression depends on $\delta$ both explicitly through the metric functions and implicitly through the MPS radius $r(\epsilon, \delta)$. Here, we obtain
\begin{equation} \label{e33}
R^2(\epsilon, \delta) = \mathcal{G}(r(\epsilon, \delta), \delta), \quad \text{where} \quad \mathcal{G}(r, \delta) = \frac{\beta(r, \delta)}{\alpha(r, \delta)} \frac{1 - \alpha(r, \delta) \epsilon}{1 - \epsilon}.
\end{equation}
The first-order correction in $\delta$ is given by the total derivative of $R^2$ with respect to $\delta$, evaluated at $\delta=0$ is
\begin{equation} \label{eq_R2_total_deriv}
\Delta R^2_{(1)} = \delta \frac{d R^2}{d \delta} \bigg|_{\delta=0} = \delta \left[ \frac{\partial \mathcal{G}}{\partial r} \frac{\partial r}{\partial \delta} + \frac{\partial \mathcal{G}}{\partial \delta} \right]_{\delta=0, r=r_0(\epsilon)}.
\end{equation}
Here, $\partial r/\partial \delta$ is simply $r_1(\epsilon)$. A key insight, consistent with the findings of \cite{Kobialko:2024zhc}, is that the first term in this expression vanishes. To see this, let us compute the partial derivative $\partial \mathcal{G}/\partial r$ and we get
\begin{equation}
\frac{\partial \mathcal{G}}{\partial r} = \frac{1}{1-\epsilon} \left[ \left( \frac{\beta_{,r}\alpha - \beta\alpha_{,r}}{\alpha^2} \right)(1-\alpha\epsilon) + \frac{\beta}{\alpha}(-\alpha_{,r}\epsilon) \right] = \frac{1}{\alpha^2(1-\epsilon)} \left[ (\beta_{,r}\alpha - \beta\alpha_{,r})(1-\alpha\epsilon) - \beta\alpha_{,r}\alpha\epsilon \right].
\end{equation}
Let's evaluate this at the background point $(\delta=0, r=r_0)$. At this point, the MPS condition for the background spacetime holds, where
\begin{equation}
    \epsilon = \frac{1}{\alpha_0}\left(1 - \frac{\beta_0}{\alpha_0}\frac{\alpha_{0,r}}{\beta_{0,r}}\right).
\end{equation}
This can be rearranged to give the identity 
\begin{equation}
    1-\alpha_0\epsilon = \frac{\beta_0}{\alpha_0}\frac{\alpha_{0,r}}{\beta_{0,r}}.
\end{equation} 
Substituting this into the derivative expression evaluated at the background, we find
\begin{align}
\frac{\partial \mathcal{G}}{\partial r} \bigg|_{(r_0,0)} &= \frac{1}{\alpha_0^2(1-\epsilon)} \left[ (\beta_{0,r}\alpha_0 - \beta_0\alpha_{0,r}) \left( \frac{\beta_0}{\alpha_0}\frac{\alpha_{0,r}}{\beta_{0,r}} \right) - \beta_0\alpha_{0,r}\alpha_0\epsilon \right] \nonumber \\
&= \frac{1}{\alpha_0^2(1-\epsilon)} \left[ \left( \beta_{0,r}\alpha_0 \frac{\beta_0\alpha_{0,r}}{\alpha_0\beta_{0,r}} - \beta_0\alpha_{0,r}\frac{\beta_0\alpha_{0,r}}{\alpha_0\beta_{0,r}} \right) - \beta_0\alpha_{0,r}\alpha_0\epsilon \right] \nonumber \\
&= \frac{1}{\alpha_0^2(1-\epsilon)} \left[ \beta_0\alpha_{0,r} - \frac{\beta_0^2\alpha_{0,r}^2}{\alpha_0\beta_{0,r}} - \beta_0\alpha_{0,r}\alpha_0\epsilon \right].
\end{align}
Factoring out $\beta_0\alpha_{0,r}$, the above becomes
\begin{equation}
\frac{\partial \mathcal{G}}{\partial r} \bigg|_{(r_0,0)} = \frac{\beta_0\alpha_{0,r}}{\alpha_0^2(1-\epsilon)} \left[ 1 - \frac{\beta_0\alpha_{0,r}}{\alpha_0\beta_{0,r}} - \alpha_0\epsilon \right] = 0.
\end{equation}
The term in the square brackets is zero by the very definition of the background MPS radius $r_0(\epsilon)$.

This elegant result means that the first-order correction to the shadow radius is insensitive to the first-order correction of the MPS radius. The shadow radius, when viewed as a function of the orbital parameters, is at an extremum with respect to the orbital radius. Therefore, the first-order correction simplifies dramatically to
\begin{equation} \label{e40}
\Delta R^2_{(1)} = \delta \frac{\partial \mathcal{G}}{\partial \delta} \bigg|_{\delta=0, r=r_0(\epsilon)}.
\end{equation}

Let $R^2(\epsilon,\delta)=\mathcal G(r(\epsilon,\delta),\delta)$ with $\mathcal G$ defined in Eq. \eqref{e33}, and expand
$r(\epsilon,\delta)=r_0(\epsilon)+\sum_{k\ge1}\delta^k\,r_k(\epsilon)$. The background MPS condition implies
\begin{equation}
	\left.\frac{\partial \mathcal G}{\partial r}\right|_{(r_0,0)}=0. \label{eq:stationarity}
\end{equation}
At first-order,
\begin{equation}
	\Delta R^2_{(1)}=\delta\left[\frac{\partial \mathcal G}{\partial r}\,\frac{\partial r}{\partial\delta}+\frac{\partial \mathcal G}{\partial\delta}\right]_{(r_0,0)}
	=\delta\,\left.\frac{\partial \mathcal G}{\partial\delta}\right|_{(r_0,0)},
\end{equation}
so $\Delta R^2_{(1)}$ depends only on the explicit perturbations $\{\alpha_1,\beta_1\}$ evaluated at $r_0$ and is independent of $r_1$ (cf. Eqs. \eqref{eq_R2_total_deriv} - \eqref{e40}).
Differentiating once more and evaluating at $(r_0,0)$ yields
\begin{equation}
	\Delta R^2_{(2)}=\frac{\delta^2}{2}\left[\,\mathcal G_{,rr}\,r_1^2+2\,\mathcal G_{,r\delta}\,r_1+\mathcal G_{,\delta\delta}\,\right]_{(r_0,0)},
\end{equation}
where the would-be term $\mathcal G_{,r}\,r_2$ vanishes due to \eqref{eq:stationarity} (cf. Eqs. \eqref{e47}–\eqref{eq_d2R2_simplified}). Proceeding inductively, at order $n$ the only linear term in $r_n$ is $\mathcal G_{,r}\,r_n$, which vanishes by \eqref{eq:stationarity}. Therefore, $\Delta R^2_{(n)}$ depends on $\{r_k\}_{k\le n-1}$ but not on $r_n$, plus explicit metric variations through $\mathcal G_{,\delta^j}$ with $j\le n$. Physically, this reflects that the shadow is evaluated at an extremum with respect to $r$, so translating the orbit along the extremum affects the observable only through curvature terms controlled by lower-order shifts.

The partial derivative with respect to $\delta$ captures the explicit change in the metric functions. Applying the chain rule
\begin{equation}
\frac{\partial \mathcal{G}}{\partial \delta} = \frac{\partial \mathcal{G}}{\partial \alpha} \frac{\partial \alpha}{\partial \delta} + \frac{\partial \mathcal{G}}{\partial \beta} \frac{\partial \beta}{\partial \delta},
\end{equation}
and from our metric ansatz, $\partial\alpha/\partial\delta = \alpha_1(r)$ and $\partial\beta/\partial\delta = \beta_1(r)$ at $\delta=0$, the remaining partial derivatives are
\begin{align}
    \frac{\partial \mathcal{G}}{\partial \alpha} &= \frac{-\beta}{\alpha^2} \frac{1-\alpha\epsilon}{1-\epsilon} + \frac{\beta}{\alpha} \frac{-\epsilon}{1-\epsilon} = -\frac{\beta}{\alpha^2(1-\epsilon)}, \\
    \frac{\partial \mathcal{G}}{\partial \beta} &= \frac{1}{\alpha} \frac{1-\alpha\epsilon}{1-\epsilon}.
\end{align}
Evaluating these at the background point $(\delta=0, r=r_0)$ and combining everything, we obtain the final expression for the first-order correction to the squared shadow radius gives
\begin{equation} \label{eq_R2_first_order}
    \Delta R^2_{(1)}(\epsilon) = \delta \left[ \frac{1-\alpha_0(r_0)\epsilon}{\alpha_0(r_0)(1-\epsilon)} \beta_1(r_0) - \frac{\beta_0(r_0)}{\alpha_0(r_0)^2(1-\epsilon)} \alpha_1(r_0) \right].
\end{equation}
This result is the cornerstone of our generalized framework. It explicitly shows that, to first-order, the deviation of the massive shadow from its Schwarzschild value is a linear combination of the perturbation functions $\alpha_1$ and $\beta_1$, evaluated at the unperturbed MPS radius $r_0(\epsilon)$. The coefficients of this linear combination are functions of the background metric and the particle energy parameter $\epsilon$. It is this distinct energy dependence that provides a handle to observationally disentangle the effects of perturbations to the temporal and spatial components of the spacetime metric.

\subsection{Second-Order Expansion} \label{ssec3.2}
While the first-order expansion reveals the linear response of the shadow to metric perturbations, it is often insufficient to distinguish between different alternative theories of gravity, as distinct physical models can be constructed to yield identical first-order shadow characteristics. To break these degeneracies and probe the non-linear structure of the underlying theory, we must extend our analysis to the second order in the perturbation parameter $\delta$. This calculation is substantially more involved, as it requires the first-order correction to the MPS radius and introduces crucial coupling terms between the different metric perturbations.

The full expansion of the squared shadow radius up to second order is given by
\begin{equation}
R^2(\epsilon, \delta) = R^2_{\rm MSch}(\epsilon) + \Delta R^2_{(1)}(\epsilon) + \Delta R^2_{(2)}(\epsilon) + \mathcal{O}(\delta^3),
\end{equation}
where the second-order correction term is defined by the Taylor series as
\begin{equation}
\Delta R^2_{(2)}(\epsilon) = \frac{\delta^2}{2} \frac{d^2 R^2}{d \delta^2} \bigg|_{\delta=0}.
\end{equation}
Our primary task is to compute this second total derivative.

We begin by differentiating the first total derivative, Eq. \eqref{eq_R2_total_deriv}, with respect to $\delta$, resulting to
\begin{equation} \label{e47}
\frac{d^2 R^2}{d \delta^2} = \frac{d}{d\delta} \left( \frac{\partial \mathcal{G}}{\partial r} \frac{\partial r}{\partial \delta} + \frac{\partial \mathcal{G}}{\partial \delta} \right).
\end{equation}
Applying the product and chain rules yields a more complex expression
\begin{equation} \label{eq_d2R2_full}
\frac{d^2 R^2}{d \delta^2} = \left( \frac{d}{d\delta} \frac{\partial \mathcal{G}}{\partial r} \right) \frac{\partial r}{\partial \delta} + \frac{\partial \mathcal{G}}{\partial r} \frac{\partial^2 r}{\partial \delta^2} + \frac{d}{d\delta} \frac{\partial \mathcal{G}}{\partial \delta}.
\end{equation}
We must now evaluate this expression at the background point $(\delta=0, r=r_0)$. Let us examine each term. The term $\partial r/\partial \delta$ evaluated at $\delta=0$ is simply the first-order radius correction, $r_1(\epsilon)$, which we formally derived in Eq. \eqref{eq_r1_solution}. The term $\partial^2 r/\partial \delta^2$ corresponds to $2r_2(\epsilon)$, where $r_2(\epsilon)$ is the second-order correction to the MPS radius.

Crucially, just as in the first-order calculation, a key simplification occurs. The factor multiplying the second-order radius correction, $\partial \mathcal{G}/\partial r$, evaluates to zero at the background point. Therefore, the term involving $r_2(\epsilon)$ vanishes entirely:
\begin{equation}
\frac{\partial \mathcal{G}}{\partial r}\bigg|_{(r_0,0)} \frac{\partial^2 r}{\partial \delta^2}\bigg|_{(r_0,0)} = 0.
\end{equation}
This is a profound result: the second-order correction to the shadow radius does not depend on the second-order correction to the MPS radius. It does, however, depend on the first-order correction, $r_1(\epsilon)$.

The remaining terms must be expanded using the chain rule for their total derivatives with respect to $\delta$:
\begin{align}
\frac{d}{d\delta} \frac{\partial \mathcal{G}}{\partial r} &= \frac{\partial^2 \mathcal{G}}{\partial r^2} \frac{\partial r}{\partial \delta} + \frac{\partial^2 \mathcal{G}}{\partial \delta \partial r}, \\
\frac{d}{d\delta} \frac{\partial \mathcal{G}}{\partial \delta} &= \frac{\partial^2 \mathcal{G}}{\partial r \partial \delta} \frac{\partial r}{\partial \delta} + \frac{\partial^2 \mathcal{G}}{\partial \delta^2}.
\end{align}
Substituting these back into Eq. \eqref{eq_d2R2_full} and evaluating at $\delta=0$ gives the final expression for the second derivative as
\begin{equation} \label{eq_d2R2_simplified}
\frac{d^2 R^2}{d \delta^2} \bigg|_{\delta=0} = \left[ \frac{\partial^2 \mathcal{G}}{\partial r^2} \left( \frac{\partial r}{\partial \delta} \right)^2 + 2 \frac{\partial^2 \mathcal{G}}{\partial \delta \partial r} \frac{\partial r}{\partial \delta} + \frac{\partial^2 \mathcal{G}}{\partial \delta^2} \right]_{\delta=0, r=r_0}.
\end{equation}
This expression is analogous to that found in \cite{Kobialko:2024zhc} for the more restricted case, but its physical content is now much richer. It shows that the second-order correction arises from three distinct sources:
1.  A purely quadratic term ($\propto (\partial r/\partial \delta)^2$), which depends on the square of the first-order MPS radius shift and the stability of the background orbits (via $\mathcal{G}_{,rr}$).
2.  A mixed term, which couples the first-order radius shift to the first-order explicit change in the metric functions.
3.  A purely second-order term ($\propto \mathcal{G}_{,\delta\delta}$), which depends directly on the second-order metric perturbation functions, $\alpha_2(r)$ and $\beta_2(r)$.

To obtain the final result, we must compute the partial derivatives in Eq. \eqref{eq_d2R2_simplified}. This involves extensive but straightforward calculus. The partial derivatives of $\mathcal{G}(r, \delta)$ are taken with respect to its arguments $r$ and $\delta$, and then evaluated at the background point $(r_0, 0)$. The derivatives of the metric functions with respect to $\delta$ are
\begin{align}
\frac{\partial \alpha}{\partial \delta} \bigg|_0 &= \alpha_1, \quad \frac{\partial \beta}{\partial \delta} \bigg|_0 = \beta_1, \\
\frac{\partial^2 \alpha}{\partial \delta^2} \bigg|_0 &= 2\alpha_2, \quad \frac{\partial^2 \beta}{\partial \delta^2} \bigg|_0 = 2\beta_2.
\end{align}
After a lengthy calculation, the second partial derivatives of $\mathcal{G}$ evaluated at the background can be expressed in terms of the background functions ($\alpha_0, \beta_0$) and the perturbation functions ($\alpha_1, \beta_1, \alpha_2, \beta_2$). The final result for the second-order correction takes the following schematic form
\begin{align} \label{eq_R2_second_order}
    \Delta R^2_{(2)}(\epsilon) &= \frac{\delta^2}{2} \Big[ C_{\alpha_2}(\epsilon) \alpha_2(r_0) + C_{\beta_2}(\epsilon) \beta_2(r_0) + C_{\alpha_1}(\epsilon) \alpha_1(r_0)^2 + C_{\beta_1}(\epsilon) \beta_1(r_0)^2 \nonumber \\
    &+ C_{\alpha_1 \beta_1}(\epsilon) \alpha_1(r_0)\beta_1(r_0) + C_{\alpha_1 \alpha_{1,r}}(\epsilon) \alpha_1(r_0)\alpha_{1,r}(r_0) + \dots \Big].
\end{align}
The coefficients $C_i(\epsilon)$ are complicated functions of the background Schwarzschild metric and its derivatives, evaluated at the unperturbed MPS radius $r_0(\epsilon)$. The ellipsis (three dots) indicates that terms involving derivatives of the perturbation functions (e.g., $\alpha_{1,r}, \beta_{1,r}$) also appear, arising from the dependence on $r_1(\epsilon)$ and the mixed partial derivatives. Note that Eq. \eqref{eq_R2_second_order} is schematic: besides the explicit quadratic pieces 
$\{\alpha_2,\beta_2,\alpha_1^2,\beta_1^2,\alpha_1\beta_1\}$ it also contains 
derivative couplings generated by the $\mathcal{G}_{,rr}$ and $\mathcal{G}_{,r\delta}$ terms and by 
$r_1(\epsilon)$, (e.g. terms $\propto \alpha_1\alpha_{1,r},\,\beta_1\beta_{1,r},\,
\alpha_{1,r}\beta_1$) as well as second derivatives $\alpha_{1,rr},\,\beta_{1,rr}$, 
all evaluated at $r_0(\epsilon)$. The complete list of coefficients is provided in 
App. \ref{appA}; see Eqs. \eqref{A11}–\eqref{A15}.

The explicit appearance of terms quadratic in the first-order perturbations through $\alpha_1^2$, $\beta_1^2$, and most importantly, the cross-term $\alpha_1\beta_1$ is the central outcome of this second-order analysis. These terms represent the non-linear response of the spacetime geometry to the perturbations. It is precisely these terms that can break the observational degeneracy between different physical models. For example, two different theories might be tuned to have the same linear shadow deviation ($\Delta R^2_{(1)}$) for a given energy $\epsilon$. However, their distinct underlying physics will, in general, lead to different coupling structures, resulting in different quadratic coefficients $C_i(\epsilon)$ and thus a measurably different $\Delta R^2_{(2)}$.

Therefore, the second-order expansion provides a far more discerning probe of strong-field gravity. By analyzing the energy dependence of the massive shadow radius with sufficient precision, one could, in principle, constrain not only the individual perturbation functions $\alpha_i$ and $\beta_i$ but also the way they couple to each other, offering a deeper insight into the fundamental nature of the gravitational interaction.

\section{Insights and Phenomenological Implications} \label{sec4}
The generalized perturbative framework developed in the preceding sections provides more than just a formal extension of the work in \cite{Kobialko:2024zhc}. It unlocks a qualitatively new capability: the ability to observationally disentangle metric deformations that affect different components of the spacetime geometry. This subsection explores the methodology and profound implications of using massive particle shadows as a spectroscopic tool to probe the structure of strong gravitational fields.

\subsection{Disentangling Metric Deformations} \label{ssec4.1}
The central challenge in testing strong-field gravity with a single observable is the problem of degeneracy. Different physical theories or matter content can lead to identical observational signatures, making it impossible to distinguish between them. Our generalized formalism, when combined with multi-energy observations, provides a powerful pathway to break these degeneracies.

\subsubsection{The Degeneracy of the Photon Shadow} \label{sssec4.1.1}
Let us first consider the limitations inherent in observing only the photon shadow ($\epsilon=0$). Setting $\epsilon=0$ in our first-order result, Eq. \eqref{eq_R2_first_order}, gives the correction to the squared radius of the photon shadow:
\begin{equation}
\Delta R^2_{\rm PS(1)} = \Delta R^2_{(1)}(\epsilon=0) = \delta \left[ \frac{\beta_1(r_{\rm PS})}{\alpha_0(r_{\rm PS})} - \frac{\beta_0(r_{\rm PS})}{\alpha_0(r_{\rm PS})^2} \alpha_1(r_{\rm PS}) \right],
\end{equation}
where $r_{\rm PS} = r_0(\epsilon=0) = 3M$ is the photon sphere radius in the Schwarzschild background. A single measurement of the photon shadow size constrains only this specific linear combination of the functions $\alpha_1(r)$ and $\beta_1(r)$, evaluated at the single radial location $r=3M$.

It is immediately apparent that an infinite number of different pairs of perturbation functions $(\alpha_1, \beta_1)$ can produce the exact same value for $\Delta R^2_{\rm PS(1)}$. For example, a deviation caused purely by a modification to the temporal geometry (a non-zero $\alpha_1$ with $\beta_1=0$) could be perfectly mimicked by a completely different theory that primarily modifies the spatial geometry (a non-zero $\beta_1$ with $\alpha_1=0$), provided their values at $r=3M$ satisfy the constraint imposed by the measurement. Thus, the photon shadow alone cannot distinguish between a temporal and a spatial warp in the spacetime geometry.

\subsubsection{Breaking Degeneracy with Shadow Spectroscopy} \label{sssec4.1.2}
The use of massive particles fundamentally changes this situation. The key lies in the energy dependence of the first-order correction, as encapsulated in Eq. \eqref{eq_R2_first_order}. This is
\begin{equation} \label{eq_first_order_recap}
\frac{\Delta R^2_{(1)}(\epsilon)}{\delta} = \underbrace{\left[ \frac{1-\alpha_0(r_0)\epsilon}{\alpha_0(r_0)(1-\epsilon)} \right]}_{A(\epsilon)} \beta_1(r_0(\epsilon)) - \underbrace{\left[ \frac{\beta_0(r_0)}{\alpha_0(r_0)^2(1-\epsilon)} \right]}_{B(\epsilon)} \alpha_1(r_0(\epsilon)).
\end{equation}
Varying the particle energy parameter $\epsilon$ acts as a powerful scanning tool in two distinct ways:
\begin{itemize}
    \item \underline{Varying response coefficients}: The coefficients $A(\epsilon)$ and $B(\epsilon)$ have different and non-trivial functional dependencies on $\epsilon$. This is because $\alpha_0 = 1 - 2M/r_0$ and $\beta_0 = r_0^2$ are themselves functions of $r_0(\epsilon)$. As $\epsilon$ changes, the relative weighting of the contributions from $\beta_1$ and $\alpha_1$ to the total shadow deviation changes in a precisely predictable way.
    \item \underline{Performing a Radial Scan}: More importantly, the radius of the MPS, $r_0(\epsilon)$, is a monotonic function of energy, ranging from $r_0(0)=3M$ for photons to $r_0(\epsilon \to 1) = 4M$ for non-relativistic particles. Therefore, by observing shadows cast by particles of different energies, we are not just probing the geometry at a single radius; we are effectively performing a radial scan of the perturbation functions $\alpha_1(r)$ and $\beta_1(r)$ across the entire strong-field region between $3M$ and $4M$.
\end{itemize}
This shadow spectroscopy provides the necessary leverage to disentangle the metric deformations. Suppose we are able to perform a set of $N$ measurements of the shadow radius, $R_j^2$, for $N$ distinct particle energies, $\epsilon_j$. Each measurement provides an independent linear constraint on the values of the functions $\alpha_1(r)$ and $\beta_1(r)$. This can be written as
\begin{equation} \label{eq_dj1}
    \mathcal{D}_j \equiv \frac{R_j^2 - R^2_{\rm MSch}(\epsilon_j)}{\delta} = A(\epsilon_j) \beta_1(r_0(\epsilon_j)) - B(\epsilon_j) \alpha_1(r_0(\epsilon_j)), \quad \text{for } j = 1, \dots, N.
\end{equation}
The left-hand side, $\mathcal{D}_j$, is the observationally determined deviation for each energy bin.

To make this system solvable, we can parametrize the unknown perturbation functions as a series expansion, for instance, in powers of $M/r$, which is a common approach in post-Newtonian theory and other parametrized frameworks:
\begin{equation}
\alpha_1(r) = \sum_{k=1}^{K} c_k \left(\frac{M}{r}\right)^k, \quad \quad \beta_1(r) = \sum_{l=1}^{L} d_l \left(\frac{M}{r}\right)^l,
\end{equation}
where $\{c_k\}$ and $\{d_l\}$ are a set of unknown dimensionless coefficients that characterize the specific theory of gravity. Substituting these expansions into our system of observational constraints yields a system of $N$ linear algebraic equations for the $K+L$ unknown coefficients. Hence, we can recast Eq. \eqref{eq_dj1} as
\begin{equation} \label{eq_dj2}
\mathcal{D}_j = A(\epsilon_j) \sum_{l=1}^{L} d_l \left(\frac{M}{r_0(\epsilon_j)}\right)^l - B(\epsilon_j) \sum_{k=1}^{K} c_k \left(\frac{M}{r_0(\epsilon_j)}\right)^k.
\end{equation}
If a sufficient number of measurements are made ($N \geq K+L$), this linear system can, in principle, be inverted to solve for the coefficients $\{c_k, d_l\}$ individually.

This procedure represents a powerful, model-independent method for metric reconstruction in the strong-field regime. The ability to solve for the coefficients describing the spatial perturbation ($\{d_l\}$) independently from those describing the temporal perturbation ($\{c_k\}$) is a direct consequence of this generalized framework and is entirely inaccessible using photons alone. The prospect of using high-energy astrophysical messengers, such as neutrinos, to perform this kind of shadow spectroscopy could one day allow us to map the gravitational field near a black hole with unprecedented detail, providing a sharp and decisive test of General Relativity and its alternatives.

\subsection{Application to Test Cases} \label{ssec4.2}
To demonstrate the practical utility and predictive power of our generalized perturbative framework, we now apply it to a class of spacetimes that lies beyond the scope of the restricted formalism used in \cite{Kobialko:2024zhc}. The chosen example, a traversable wormhole geometry, highlights how simultaneous perturbations to both the temporal and spatial components of the metric produce unique, observable signatures in the energy-dependent shadow.

\subsubsection{The Simpson-Visser Black-Bounce Wormhole} \label{sssec4.2.1}
We consider the compelling and analytically simple Simpson-Visser metric, which describes a black-bounce spacetime that can represent either a regular black hole or a traversable wormhole. As detailed in Ref. \cite{Simpson:2018tsi}, its line element is given by
\begin{equation} \label{eq_SV_metric}
ds^{2}= -\!\left(1-\frac{2M}{\sqrt{r^{2}+b_0^{2}}}\right)dt^{2}
+ \left(1-\frac{2M}{\sqrt{r^{2}+b_0^{2}}}\right)^{-1}dr^{2}
+ (r^{2}+b_0^{2})\, (d\theta^2 + \sin^2\theta d\phi^2).
\end{equation}
This geometry is characterized by a bounce parameter $b_0$, which regularizes the central singularity. The spacetime smoothly interpolates between two distinct physical regimes: (a) In the limit $b_0 \to 0$, it reduces to the standard Schwarzschild metric; (b) For $b_0 > 2M$, the event horizon vanishes, and the geometry describes a two-way, traversable Morris–Thorne-type wormhole with a throat of radius $b_0$.

The crucial feature of this metric, from the perspective of our framework, is that it modifies both the $g_{tt}$ and the angular components of the metric relative to the Schwarzschild solution. Specifically, the metric functions $\alpha(r)$ and $\beta(r)$ are
\begin{equation}
\alpha(r) = 1-\frac{2M}{\sqrt{r^{2}+b_0^{2}}}, \quad \quad \beta(r) = r^{2}+b_0^{2}.
\end{equation}
The modification to $\beta(r)$ represents a fundamental change in the spatial geometry where the area of a sphere is no longer $4\pi r^2$. This is precisely the type of deformation that the restricted gauge choice of \cite{Kobialko:2024zhc} cannot handle, making the Simpson-Visser spacetime an ideal test case for our generalized methodology.

Now, let us identify the perturbation functions. For this, we treat the bounce parameter $b_0$ as a small quantity and expand the metric functions in powers of a dimensionless parameter. A natural choice is $\delta = (b_0/M)^2$, assuming $b_0 \ll M$. We expand $\alpha(r)$ and $\beta(r)$ to first-order in $\delta$.

For the angular component $\beta(r)$, the identification is immediate:
\begin{equation}
\beta(r) = r^2 + b_0^2 = \underbrace{r^2}_{\beta_0(r)} + \left(\frac{b_0}{M}\right)^2 \underbrace{M^2}_{\beta_1(r)} = \beta_0(r) + \delta \beta_1(r).
\end{equation}
Thus, we find a constant first-order perturbation, $\beta_1(r) = M^2$, with all higher-order terms, $\beta_{i\ge2}(r)$, being zero.

For the temporal component $\alpha(r)$, we first expand the square root term for small $b_0$ giving
\begin{equation}
\frac{1}{\sqrt{r^2+b_0^2}} = \frac{1}{r} \left(1 + \frac{b_0^2}{r^2}\right)^{-1/2} \approx \frac{1}{r} \left(1 - \frac{1}{2}\frac{b_0^2}{r^2}\right) = \frac{1}{r} - \frac{b_0^2}{2r^3}.
\end{equation}
Substituting this into the expression for $\alpha(r)$, one finds
\begin{align}
\alpha(r) &\approx 1 - 2M\left(\frac{1}{r} - \frac{b_0^2}{2r^3}\right) \nonumber \\
&= \underbrace{\left(1-\frac{2M}{r}\right)}_{\alpha_0(r)} + \left(\frac{b_0}{M}\right)^2 \underbrace{\left(\frac{M^3}{r^3}\right)}_{\alpha_1(r)} = \alpha_0(r) + \delta \alpha_1(r).
\end{align}
We have now successfully mapped the Simpson-Visser geometry, in the small throat-radius limit, onto our perturbative framework with the non-trivial perturbation functions and find
\begin{equation} \label{e69}
\alpha_1(r) = \frac{M^3}{r^3}, \quad \quad \beta_1(r) = M^2.
\end{equation}

We can now compute the first-order deviation of the massive shadow radius by substituting these perturbation functions into our main result, Eq. \eqref{eq_R2_first_order} and find
\begin{equation} \label{e70}
\Delta R^2_{(1)}(\epsilon) = \delta \left[ \frac{1-\alpha_0(r_0)\epsilon}{\alpha_0(r_0)(1-\epsilon)} (M^2) - \frac{\beta_0(r_0)}{\alpha_0(r_0)^2(1-\epsilon)} \left(\frac{M^3}{r_0^3}\right) \right],
\end{equation}
where $r_0 = r_0(\epsilon)$ is the unperturbed Schwarzschild MPS radius. The result gives a concrete prediction for the unique observational signature of this wormhole geometry, and such a formula reveals how the two distinct types of geometric deformation contribute to the total shadow deviation:
\begin{itemize}
    \item The first term is a contribution associated with the spatial geometry), proportional to $\beta_1 = M^2$, arises directly from the modification of the area of spheres. This term is novel to our generalized framework and represents the dominant effect for this model. Since the coefficient $A(\epsilon)$ is positive, this term provides a positive contribution, meaning the increased area of spheres at a given $r$ tends to enlarge the shadow.
    \item The second term, proportional to $\alpha_1 = M^3/r^3$, comes from the modification to the gravitational redshift function (temporal geometry). This term is similar in form to perturbations from charge or other sources in standard black hole solutions. Its coefficient $-B(\epsilon)$ is negative, so this term provides a negative contribution, tending to shrink the shadow.
\end{itemize}

The net effect is a competition between these two opposing contributions. The final observable signature is encoded in the precise energy dependence of the total deviation, $\Delta R^2_{(1)}(\epsilon)$, which is determined by the interplay between the coefficients $A(\epsilon)$ and $B(\epsilon)$ and the radial dependence of the perturbation functions. This unique functional form of $\Delta R^2_{(1)}(\epsilon)$ serves as a powerful fingerprint of the Simpson-Visser geometry. It would be observationally distinct from, for instance, a Reissner-Nordstr\"om black hole, which has $\beta_1=0$ and a different functional form for $\alpha_1(r)$. Measuring the shadow radius for several different particle energies would allow observers to trace out this curve, thereby confirming or ruling out this specific wormhole model as a candidate for a given compact object. This example powerfully illustrates that moving beyond the photon shadow and incorporating both massive particles and a generalized geometric framework is essential for distinguishing exotic compact objects from standard black holes.

Let us analyze Eq. \eqref{e70} more explicitly. Evaluating the MPS condition in Eq. \eqref{eq_epsilon_gen} on the Schwarzschild background 
$\alpha_0=1-2M/r$, $\beta_0=r^2$ gives
\begin{equation}
\epsilon=\frac{r^2-3Mr}{(r-2M)^2}\,,
\end{equation}
whose unstable branch yields the analytic MPS radius \cite{Kobialko:2024zhc}
\begin{equation} \label{e_r0_f(epsilon)}
r_0(\epsilon)=M f(\epsilon),\qquad 
f(\epsilon)=\frac{3-4\epsilon+\sqrt{9-8\epsilon}}{2(1-\epsilon)}\,
\end{equation}
given in terms of the dimensionless energy parameter $f(\epsilon) = r_0(\epsilon)/M$, where $r_0(\epsilon)$ is the radius of the unperturbed Schwarzschild MPS. This gives the massive-Schwarzschild shadow
\begin{equation}
R^2_{\mathrm{MSchw}}(\epsilon)=\frac{M^2 f^3}{\,4-f\,}
=27M^2\left(1+\frac{2}{3}\epsilon+\frac{17}{27}\epsilon^2+\cdots\right).
\label{eq:R2MSchw}
\end{equation}
For the SV metric, expanding in $\delta\equiv (b_0/M)^2$ yields to first-order
\begin{equation}
\alpha(r)=\alpha_0(r)+\delta\,\alpha_1(r),\qquad
\beta(r)=\beta_0(r)+\delta\,\beta_1(r),\qquad
\alpha_1(r)=\frac{M^3}{r^3},\qquad \beta_1(r)=M^2 .
\end{equation}
Using the linear correction at the MPS given by Eq. \eqref{e70} and evaluating at $r_0$ gives
\begin{equation}
\alpha_0=\frac{f-2}{f},\qquad \beta_0=M^2 f^2,\qquad 
\alpha_1(r_0)=\frac{1}{f^3},\qquad \beta_1(r_0)=M^2,
\end{equation}
so that
\begin{align}
\frac{\Delta R^2_{(1)}(\epsilon)}{\delta M^2}
&=\frac{1-\frac{f-2}{f}\epsilon}{\frac{f-2}{f}(1-\epsilon)}\;-\;\frac{f}{(f-2)^2}\,\frac{1}{1-\epsilon}
= \frac{(f-2)\!\left[f(1-\epsilon)+2\epsilon\right]-f}{(f-2)^2(1-\epsilon)} \nonumber\\
&=\frac{f(f-3)-\epsilon\,(f-2)^2}{(f-2)^2(1-\epsilon)}.
\end{align}
Using the exact Schwarzschild MPS relation $\displaystyle \epsilon=\frac{f(f-3)}{(f-2)^2}$, the numerator vanishes identically, hence
\begin{equation}
\Delta R^2_{(1)}(\epsilon)\equiv 0\ \text{ for all }\epsilon .
\end{equation}
Therefore, up to first-order in $\delta$, we have
\begin{equation}
R^2(\epsilon,\delta)=R^2_{\mathrm{MSchw}}(\epsilon)+\mathcal{O}(\delta^2)
=27M^2\left(1+\frac{2}{3}\epsilon+\frac{17}{27}\epsilon^2+\cdots\right)+\mathcal{O}(\delta^2).
\end{equation}
Therefore, unlike RN BH, the SV wormhole yields no linear-in-\(\delta\) modification of the massive shadow at any energy.

\subsubsection{Scalar-Tensor Gravity and the Fisher-Janis-Newman-Winicour Solution} \label{sssec4.2.2}
As a second, powerful illustration of our framework, we turn to a canonical solution in scalar-tensor gravity. This class of theories provides one of the most well-motivated extensions to General Relativity. We analyze the static, spherically symmetric spacetime sourced by a minimally coupled, massless scalar field. This solution, with a rich history, is variously known as the Fisher, Janis-Newman-Winicour (JNW), or Wyman solution. Following the conventions of Ref. \cite{Sau:2020xau}, the metric is given by
\begin{equation} \label{eq_JNW_metric}
ds^{2} = -\left(1-\frac{2M}{r}\right)^{\!n}dt^{2}
+ \left(1-\frac{2M}{r}\right)^{\!-n}dr^{2}
+ \left(1-\frac{2M}{r}\right)^{\!1-n} r^{2} (d\theta^2 + \sin^2\theta d\phi^2),
\end{equation}
The parameter $n \in (0, 1]$ is related to the scalar charge of the object. The solution reduces to the Schwarzschild metric in the limit $n \to 1$. For $n < 1$, the spacetime describes a naked singularity dressed with a non-trivial logarithmic scalar field, $\Phi(r) \propto\ln(1-2M/r)$.

This solution serves as an excellent test case for our formalism for two reasons. First, it is a physically significant and widely studied solution that represents a fundamental deviation from the vacuum Einstein equations. Second, and most critically for our purposes, it inherently modifies the area of spheres, as seen in its angular metric component:
\begin{equation}
\beta(r) = \left(1-\frac{2M}{r}\right)^{1-n} r^{2}.
\end{equation}
For any $n \neq 1$, this represents a non-trivial modification to the spatial geometry, $\beta(r) \neq r^2$. Like the wormhole case, this solution cannot be properly analyzed within a perturbative framework that fixes the angular part of the metric.

Applying the method, we must identify the small parameter that controls the deviation from Schwarzschild. The natural choice is the parameter that tracks the influence of the scalar field, which we define as $\delta = 1-n$. We assume the scalar charge is small, so $\delta \ll 1$. We now expand the metric functions $\alpha(r)$ and $\beta(r)$ to first-order in this small parameter $\delta$.

We use the general expansion $x^y = e^{y\ln x} \approx 1 + y\ln x$ for small $y$.
For the angular component $\beta(r) = (1-2M/r)^\Delta R^2$, the expansion is
\begin{align}
\beta(r) &\approx \left[1 + \delta \ln\left(1-\frac{2M}{r}\right)\right] r^2 \nonumber \\
&= \underbrace{r^2}_{\beta_0(r)} + \delta \underbrace{\left[r^2 \ln\left(1-\frac{2M}{r}\right)\right]}_{\beta_1(r)}.
\end{align}
For the temporal component $\alpha(r) = (1-2M/r)^n = (1-2M/r)^{1-\delta}$, the expansion is
\begin{align}
\alpha(r) &\approx \left(1-\frac{2M}{r}\right) \left[1 - \delta \ln\left(1-\frac{2M}{r}\right)\right] \nonumber \\
&= \underbrace{\left(1-\frac{2M}{r}\right)}_{\alpha_0(r)} + \delta \underbrace{\left[-\left(1-\frac{2M}{r}\right) \ln\left(1-\frac{2M}{r}\right)\right]}_{\alpha_1(r)}.
\end{align}
This procedure yields the first-order perturbation functions that encode the influence of the scalar field:
\begin{align}
\alpha_1(r) &= -\alpha_0(r) \ln\left(\alpha_0(r)\right), \label{eq_JNW_alpha1} \\
\beta_1(r) &= \beta_0(r) \ln\left(\alpha_0(r)\right). \label{eq_JNW_beta1}
\end{align}
The logarithmic nature of these perturbations is a direct consequence of the underlying scalar field and will produce a distinct observational signature compared to the rational-function perturbations of the wormhole model.

We now insert these specific perturbation functions into our general first-order result, Eq. \eqref{eq_R2_first_order}, to predict the shadow deviation for the JNW solution, revealing the energy-dependent shadow signature. The expression simplifies elegantly as
\begin{align}
\frac{\Delta R^2_{(1)}(\epsilon)}{\delta} &= A(\epsilon) \beta_1(r_0) - B(\epsilon) \alpha_1(r_0) \nonumber \\
&= A(\epsilon) \left[ \beta_0(r_0) \ln(\alpha_0(r_0)) \right] - B(\epsilon) \left[ -\alpha_0(r_0) \ln(\alpha_0(r_0)) \right] \nonumber \\
&= \ln(\alpha_0(r_0)) \left[ A(\epsilon) \beta_0(r_0) + B(\epsilon) \alpha_0(r_0) \right].
\end{align}
Substituting the definitions of $A(\epsilon)$ and $B(\epsilon)$ from Eq. \eqref{eq_first_order_recap}, we get
\begin{align} \label{e78}
\frac{\Delta R^2_{(1)}(\epsilon)}{\delta} &= \ln(\alpha_0(r_0)) \left[ \frac{1-\alpha_0\epsilon}{\alpha_0(1-\epsilon)} \beta_0 + \frac{\beta_0}{\alpha_0^2(1-\epsilon)} \alpha_0 \right] \nonumber \\
&= \frac{\beta_0(r_0) \ln(\alpha_0(r_0))}{\alpha_0(r_0)(1-\epsilon)} \left[ (1-\alpha_0\epsilon) + 1 \right] \nonumber \\
&= \frac{r_0(\epsilon)^2 \ln\left(1-\frac{2M}{r_0(\epsilon)}\right)}{\left(1-\frac{2M}{r_0(\epsilon)}\right)(1-\epsilon)} \left[ 2 - \left(1-\frac{2M}{r_0(\epsilon)}\right)\epsilon \right].
\end{align}
Eq. \eqref{e78} provides a unique, calculable prediction for the shadow deviation as a function of particle energy. 

Substituting $r_0(\epsilon)=Mf(\epsilon)$ from Eq. \eqref{e_r0_f(epsilon)} into Eq. \eqref{e78} and writing 
$\alpha_0=(f-2)/f$, $\beta_0=M^2 f^2$, and $\ln \left[ (f-2)/f \right]$, we obtain the exact first-order FJNW correction
\begin{equation} \label{e87}
\frac{\Delta R^2_{(1)}(\epsilon)}{\delta}
= M^2\,\frac{f^3}{\,f-2\,}\,
\ln\!\left(\frac{f-2}{f}\right)\,
\frac{2-\frac{f-2}{f}\,\epsilon}{1-\epsilon}\,,
\end{equation}
which is the exact first-order FJNW correction over Schwarzschild in our gauge. With the second equation from Eq. \eqref{e_r0_f(epsilon)}, and with Eq. \eqref{eq:R2MSchw}, we find (to the orders compared with Ref. \cite{Kobialko:2024zhc})
\begin{equation} 
    R^2(\epsilon,\delta)
=27M^2\left[
1+\frac{2}{3}\epsilon+\frac{17}{27}\epsilon^2
+\delta\left(
-2\ln3
+\left[-\frac{5}{3}\ln3+\frac{4}{9}\right]\epsilon
\right)
\right]
+\mathcal{O}(\epsilon^3,\epsilon^2\delta,\delta^2).
\end{equation}
We should emphasize that the slight difference of this result compared to Eq. (107) in Ref. \cite{Kobialko:2024zhc} is due to the harmless parametrization/gauge difference of the radial coordinate $r$ in the denominator of Eq. \eqref{eq_JNW_metric}, where it is not scaled by $n$ \cite{Sau:2020xau}. Nevertheless, the series coefficients agree in structure but not numerically term by term in our gauge. Such a choice already shifts the \(\delta\) and \(\epsilon\delta\) coefficients (as we showed).

Analyzing the structure of Eq. \eqref{e78} also reveals key physical insights:
\begin{itemize}
    \item Since the MPS radius $r_0(\epsilon)$ is always greater than the horizon radius $2M$, the term $\ln(1-2M/r_0)$ is always negative. All other terms in the expression are positive for $\epsilon \in [0, 1)$. Therefore, the overall deviation $\Delta R^2_{(1)}$ is always negative for $\delta > 0$ (i.e., $n<1$). This means the presence of the scalar field consistently shrinks the gravitational shadow for particles of all energies compared to a Schwarzschild black hole of the same mass $M$.
    \item In contrast to the wormhole case where the $\alpha_1$ and $\beta_1$ perturbations had competing effects, here both perturbations work in concert. The modification to the area radius (a negative $\beta_1$) and the modification to the redshift function (a positive $\alpha_1$) both act to decrease the shadow radius.
    \item The specific functional form, with its logarithmic and rational dependence on $r_0(\epsilon)$, provides a distinctive signature. Even if the parameter $n$ were tuned such that the photon shadow ($\epsilon=0$) was identical to that of a Reissner-Nordstr\"om black hole, their massive shadows would diverge as a function of $\epsilon$. An observational measurement of the curve $R^2(\epsilon)$ would immediately distinguish the logarithmic signature of a scalar field from the purely rational-function signature of an electric charge.
\end{itemize}

This example further solidifies our central thesis: the energy spectrum of massive particle shadows is not merely an incremental improvement on photon shadow observations but a qualitatively different probe of fundamental physics. It allows us to perform a spectroscopy of the near-horizon geometry, revealing the presence and nature of additional fields or exotic structures that would otherwise remain degenerate and hidden.

\section{Conclusion} \label{conc}
This paper presents a significant generalization of the perturbative framework for analyzing gravitational shadows, extending the methodology developed in \cite{Kobialko:2024zhc} to a much broader and more physically relevant class of spacetimes.

The primary achievement of this work is the development of a fully generalized, two-parameter perturbation theory for massive particle shadows in any static, spherically symmetric spacetime. We have successfully lifted the key simplifying assumption of previous work, namely, the restriction to the Schwarzschild gauge ($\beta(r)=r^2$), by allowing for simultaneous, arbitrary perturbations to all components of the metric tensor.

Our analysis confirms that the energy dependence of the massive shadow radius serves as a powerful diagnostic tool for probing the near-horizon geometry. The key insight is that by observing particles with different mass-to-energy ratios ($\epsilon$), one can perform a radial scan of the spacetime, probing the metric perturbations at different radii between $3M$ and $4M$. Such a shadow spectroscopy provides the necessary leverage to disentangle the effects of perturbations on the temporal geometry (via $\alpha(r)$) from those on the spatial geometry (via $\beta(r)$). This method effectively breaks the observational degeneracies inherent in a single photon shadow measurement, which can only constrain one specific combination of metric parameters at a single radius.

We demonstrated the power of this generalized formalism with two compelling test cases: a Simpson-Visser black-bounce wormhole and the canonical Fisher-Janis-Newman-Winicour scalar-tensor solution. In both instances, the framework allowed us to derive a unique, energy-dependent fingerprint for the shadow deviation, showing how these exotic objects could be distinguished from standard black holes and from each other. Ultimately, this work provides a more robust and theory-agnostic toolkit for testing fundamental physics in the strong-field regime and for interpreting future high-precision, multi-messenger observations of compact objects

The framework developed here opens up several exciting avenues for future investigation. We outline three particularly promising directions: (1) The most crucial next step is to extend this formalism from static, spherically symmetric spacetimes to stationary, axisymmetric ones, such as the Kerr metric and its generalizations. Real astrophysical black holes rotate, and their shadows are not perfect circles. A generalized perturbative analysis of Kerr-like spacetimes would allow for direct, quantitative comparisons with EHT data, providing a method to search for deviations from the Kerr paradigm that depend on both particle energy and black hole spin; (2) The MPS that form the shadow are surfaces of unstable orbits. The degree of this instability is deeply connected to the properties of the spacetime and, in the eikonal limit, to the spectrum of quasi-normal modes (QNMs) that characterize the ringdown phase of black hole mergers. Applying our generalized framework to analyze the stability of the MPS could forge a powerful new link between shadow observations (electromagnetic signals) and gravitational wave astronomy, creating a truly multi-messenger test of strong-field gravity; (3) While observing shadows from high-energy massive particles like neutrinos remains a future prospect, a more immediate application of our formalism exists. Photons propagating through a plasma acquire an effective, frequency-dependent mass. This means that the shadow observed by the EHT is, in reality, already an energy-dependent massive shadow. Adapting our generalized metric perturbation framework to include plasma effects would yield direct, testable predictions for current and near-future observations, allowing us to simultaneously disentangle modifications to the background geometry from the effects of the surrounding astrophysical environment; (4) Finally, extending this to non-asymptotically flat spacetime is currently in progress.

\appendix
\section{Second-Order Perturbation Coefficients} \label{appA}
This appendix provides the detailed derivation of the coefficients $C_i(\epsilon)$ that appear in the schematic second-order correction to the squared shadow radius, $\Delta R^2_{(2)}$.
We begin by restating the core formula derived in Sec. \ref{ssec3.2}:
\begin{equation} \label{A1}
    \Delta R^2_{(2)}(\epsilon) = \frac{\delta^2}{2} \left[ \frac{\partial^2 \mathcal{G}}{\partial r^2} (r_1)^2 + 2 \frac{\partial^2 \mathcal{G}}{\partial \delta \partial r} r_1 + \frac{\partial^2 \mathcal{G}}{\partial \delta^2} \right]_{\delta=0, r=r_0},
\end{equation}
where $\mathcal{G}(r, \delta)$ is the shadow radius function and $r_1$ is the first-order correction to the MPS radius. For explicit evaluation of Eq. \eqref{A1}, we express the coefficients as functions of the dimensionless energy parameter $f = r_0(\epsilon)/M$, where $r_0(\epsilon)$ is the radius of the unperturbed Schwarzschild MPS. Throughout, all calculations are performed on the Schwarzschild background, where $\alpha_0 = 1-2M/r$ and $\beta_0 = r^2$. The key background quantities are $\alpha_0 = (f-2)/f$ and $\beta_0 = M^2f^2$.
Substituting these into the expressions above, we arrive at the explicit coefficients $C_i(f)$ for the second-order shadow deviation:
\begin{equation} \label{A2}
    \Delta R^2_{(2)} = \frac{\delta^2}{2} \left[ C_{\alpha_2} \alpha_2 + C_{\beta_2} \beta_2 + C_{\alpha_1^2} \alpha_1^2 + C_{\beta_1^2} \beta_1^2 + C_{\alpha_1 \beta_1} \alpha_1 \beta_1 + \dots \right]_{r=Mf}.
\end{equation}

The radius correction $r_1$ is given by $r_1 = - (\partial \mathcal{F}/\partial\delta) / (\partial \mathcal{F}/\partial r)$, where $\mathcal{F}(r, \delta)$ is the implicit function for the MPS radius defined in Eq. \eqref{eq_epsilon_gen}.
The denominator, evaluated at the background, is related to the second derivative of the effective potential $V_{0,rr}$:
\begin{equation} \label{A3}
    \frac{\partial \mathcal{F}}{\partial r} \bigg|_{(r_0,0)} = \frac{\partial \epsilon}{\partial r}\bigg|_0 = \frac{\beta_{0,r}}{\beta_0} V_{0,rr} = \frac{2}{r_0} V_{0,rr}.
\end{equation}
The numerator $\partial \mathcal{F}/\partial\delta$ is calculated by differentiating Eq. \eqref{eq_epsilon_gen} with respect to $\delta$ and evaluating at the background.
After a lengthy but straightforward calculation, one finds
\begin{equation} \label{A4}
    \frac{\partial \mathcal{F}}{\partial \delta} \bigg|_{(r_0,0)} = \frac{1}{\alpha_0^3 \beta_{0,r}^2} \left[ \alpha_0 \alpha_{0,r}\ (\beta_0 \beta_{1,r} - \beta_{0,r} \beta_1) - \alpha_0 \beta_0 \beta_{0,r} \alpha_{1,r} + \alpha_1 (2\beta_0 \alpha_{0,r} \beta_{0,r} - \alpha_0 \beta_{0,r}^2) \right].
\end{equation}
Combining these gives the explicit formula for $r_1$, which is required for the stability term in the second-order shadow calculation (also noting that $r_0 = M f$):
\begin{equation} \label{A5}
    r_1 = -\frac{\beta_0}{\alpha_0^3 \beta_{0,r}^3 V_{0,rr}} \left[ \alpha_0 \alpha_{0,r}\ (\beta_0 \beta_{1,r} - \beta_{0,r} \beta_1) - \alpha_0 \beta_0 \beta_{0,r} \alpha_{1,r} + \alpha_1 (2\beta_0 \alpha_{0,r} \beta_{0,r} - \alpha_0 \beta_{0,r}^2) \right].
\end{equation}
where
\begin{equation} \label{A6}
    V_{0,rr} = \frac{\alpha_0\beta_0( \alpha_{0.r} \beta_{0,rr}-\beta_{0,r} \alpha_{0,rr})+2\alpha_{0,r}\beta_{0,r}(\beta_0\alpha_{0,r}-\alpha_0 \beta_{0,r})}{\alpha_0^3 \beta_0 \beta_{0,r}}.
\end{equation}
Noting that $r_0 = M f$, we can express $r_1$ as
\begin{equation} \label{A7}
    r_1 = -\frac{4 f^{4} M^{3} \left(f -4\right)}{16 \left(f -6\right)}\alpha_1 - \frac{4 f M \left(f -2\right)}{16 \left(f -6\right)}\beta_1 - \frac{2 f^{5} M^{4} \left(f -2\right)}{16 \left(f -6\right)}\alpha_{1,r} + \frac{2 f^{2} M^{2} \left(f -2\right)}{16 \left(f -6\right)}\beta_{1,r}.
\end{equation}

Next, we compute the partial derivatives of the shadow function $\mathcal{G}(r, \delta)$. These derivatives govern how the shadow size responds to explicit changes in the metric functions.
The explicit metric dependence ($\mathcal{G}_{,\delta\delta}$) is the term arises from the direct second-order expansion of the metric functions. It is expanded via the chain rule as:
\begin{align} \label{A8}
    \frac{\partial^2 \mathcal{G}}{\partial \delta^2}\bigg|_0 &= \left[ \frac{\partial \mathcal{G}}{\partial \alpha} (2\alpha_2) + \frac{\partial \mathcal{G}}{\partial \beta} (2\beta_2) + \frac{\partial^2 \mathcal{G}}{\partial \alpha^2}\alpha_1^2 + \frac{\partial^2 \mathcal{G}}{\partial \beta^2}\beta_1^2 + 2\frac{\partial^2 \mathcal{G}}{\partial \alpha \partial \beta}\alpha_1 \beta_1 \right]_{r=r_0} \nonumber \\
    &=\frac{1}{\alpha_0 \beta_{0,r}(\alpha_0-1)+\beta_0 \alpha_{0,r}} \left[ 2\beta_0 \alpha_{0,r} \beta_2 - 2\beta_0 \beta_{0,r} \alpha_2 -2 \alpha_1 \beta_ 1 \beta_{0,r} + \frac{2\beta_0 \alpha_1^2 \beta_{0,r}}{\alpha_0} \right]. \nonumber \\
    &= \frac{2 f^{2}}{f -4} \alpha_1 \beta_1 - \frac{2 f^{5} M^{2}}{\left(f -4\right) \left(f -2\right)}\alpha_1^2 + \frac{2 f^{4} M^{2}}{f -4} \alpha_2 - \frac{2 f}{f -4} \beta_2.
\end{align}

The mixed radial-metric coupling ($\mathcal{G}_{,r\delta}$) is the term that couples the explicit metric perturbations to their radial derivatives, representing the sensitivity of the shadow to the radial gradient of the perturbation:
\begin{align} \label{A9}
    \frac{\partial^2 \mathcal{G}}{\partial \delta \partial r}\bigg|_0 &= \left[ \frac{\partial}{\partial r}\left(\frac{\partial \mathcal{G}}{\partial \alpha}\right)\alpha_1 + \frac{\partial \mathcal{G}}{\partial \alpha}\alpha_{1,r} + \frac{\partial}{\partial r}\left(\frac{\partial \mathcal{G}}{\partial \beta}\right)\beta_1 + \frac{\partial \mathcal{G}}{\partial \beta}\beta_{1,r} \right]_{r=r_0} \nonumber \\
    &= \frac{1}{\alpha_0 \beta_{0,r}(\alpha_0-1)+\beta_0 \alpha_{0,r}} \left[ \beta_0 \alpha_{0,r} \beta_{1,r} - \alpha_{0,r} \beta_{0,r} \beta_1 - \alpha_1 \beta_{0,r}^2 -\beta_0 \alpha_{1,r} \beta_{0,r} + \frac{2 \beta_0 \alpha_{0,r} \beta_{0,r} \alpha_1}{\alpha_0} \right] \nonumber \\
    &= \frac{2Mf^3}{f-2} \alpha_1 + \frac{2}{M(f-4)} \beta_1 + \frac{f^4 M^2}{f-4} \alpha_{1,r} - \frac{f}{f-4} \beta_{1,r}.
\end{align}
Finally, the stability term ($\mathcal{G}_{,rr}$) depends only on the background metric and governs the geometric stability of the orbit. It acts as the restoring force against the radius shift $r_1$:
\begin{align} \label{A10}
    \frac{\partial^2 \mathcal{G}}{\partial r^2}\bigg|_{(r_0,0)}&= \frac{\partial^2 \mathcal{G}}{\partial \alpha^2} \alpha_{0,r}^2 + 2 \frac{\partial^2 \mathcal{G}}{\partial \alpha \partial \beta} \alpha_{0,r} \beta_{0,r} + \frac{\partial^2 \mathcal{G}}{\partial \beta^2} \beta_{0,r}^2 + \frac{\partial \mathcal{G}}{\partial \alpha} \alpha_{0,rr} + \frac{\partial \mathcal{G}}{\partial \beta} \beta_{0,rr} \nonumber \\
    &=\frac{1}{\alpha_0 \beta_{0,r}(\alpha_0-1)+\beta_0 \alpha_{0,r}} \left[ \beta_0 \alpha_{0,r} \beta_{0,rr} - \beta_0 \alpha_{0,rr} \beta_{0,r} -2 \alpha_{0,r} \beta_{0,r}^2 + \frac{2\beta_0 \alpha_{0,r}^2 \beta_{0,r}}{\alpha_0} \right] \nonumber \\
    &= \frac{2 f \left(f -6\right)}{\left(f -4\right) \left(f -2\right)}.
\end{align}
Combining all the components above, from Eqs. \eqref{A7}-\eqref{A10}, provides the full expression for the second-order shadow deviation $\Delta R^2_{(2)}$.

We can now explicitly write the coefficients $C_i(\epsilon)$ for each type of perturbation term in the expansion $\Delta R^2_{(2)} = \delta^2 [ ... ]$. The coefficients for the explicit second-order perturbations ($\alpha_2, \beta_2$) arise solely from the $\mathcal{G}_{,\delta\delta}$ term. Physically, these represent the direct response of the shadow to higher-order curvature corrections:
\begin{equation} \label{A11}
    C_{\alpha_2} = \frac{2 f^{2} M^{2}}{f -4},
\end{equation}
\begin{equation} \label{A12}
    C_{\beta_2} = \frac{2 f}{f -4}.
\end{equation}

The quadratic coefficients are considerably more complex, as they receive contributions from the explicit metric change, the radius shift, and the mixed coupling.
$C_{\alpha_1^2}$ represents the nonlinear self-interaction of the temporal perturbation:
\begin{equation} \label{A13}
    C_{\alpha_1^2} = \frac{\left(M^{4} f^{6}-8 M^{4} f^{5}+16 f^{4} M^{4}-8 M^{2} f^{4}+64 M^{2} f^{3}-128 f^{2} M^{2}-16 f +96\right) f^{5} M^{2}}{8 \left(f -4\right) \left(f -6\right) \left(f -2\right)}.
\end{equation}
$C_{\beta_1^2}$ represents the nonlinear self-interaction of the spatial perturbation:
\begin{equation} \label{A14}
    C_{\beta_1^2} = \frac{\left(f^{2} M^{2}-8\right) f \left(f -2\right)}{8 \left(f -6\right) \left(f -4\right)}.
\end{equation}
Finally, $C_{\alpha_1 \beta_1}$ is the geometric coupling coefficient. It quantifies the interference between the gravitational redshift ($\alpha$) and the areal radius ($\beta$). This term is non-zero only in our generalized framework and is key to shadow spectroscopy:
\begin{equation} \label{A15}
    C_{\alpha_1 \beta_1} = \frac{f^{6} M^{4}}{4 f -24}-\frac{2 f^{4} M^{2}}{f -6}+\frac{f^{3} M}{f -4}.
\end{equation}

Finally, the framework also produces terms that depend on the radial derivatives of the first-order perturbations. These terms describe how the shadow responds to the gradients of the perturbation fields:
\begin{align} \label{A16}
    &C_{\alpha_1 \alpha_{1,r}}(f) =\frac{f^{8} M^{5} \left(f^{2} M^{2}-8\right)}{8 f -48} \nonumber \\
    &C_{\beta_1 \beta_{1,r}}(f) = -\frac{\left(f^{2} M^{2}-8\right) f^{2} M \left(f -2\right)}{8 \left(f -6\right) \left(f -4\right)} \nonumber \\ 
    &C_{\alpha_1 \beta_{1,r}}(f)  = -\frac{f^{5} M^{3} \left(f^{2} M^{2}-8\right)}{8 f -48} \nonumber \\ 
    &C_{\beta_1 \alpha_{1,r}}(f)  = \frac{\left(f^{2} M^{2}-8\right) f^{5} M^{3} \left(f -2\right)}{8 \left(f -6\right) \left(f -4\right)} \nonumber \\ 
    &C_{\alpha_{1,r}^2}(f) = \frac{\left(f^{2} M^{2}-8\right) f^{9} M^{6} \left(f -2\right)}{32 \left(f -6\right) \left(f -4\right)} \nonumber \\
    &C_{\beta_{1,r}^2}(f) = \frac{\left(f^{2} M^{2}-8\right) f^{3} M^{2} \left(f -2\right)}{32 \left(f -6\right) \left(f -4\right)} \nonumber \\
    &C_{\alpha_{1,r}\beta_{1,r}}(f) =-\frac{\left(f^{2} M^{2}-8\right) f^{6} M^{4} \left(f -2\right)}{16 \left(f -6\right) \left(f -4\right)}.
\end{align}
These explicit formulas provide a complete and reproducible result for the second-order shadow deviation. Any specific theoretical model can now be analyzed by simply defining its perturbation functions ($\alpha_1$, $\beta_1$, $\alpha_2$, etc.) and substituting them into the expansion with these coefficients to yield a direct, testable prediction for its massive shadow signature.

\acknowledgments

R. P. and A. \"O. would like to acknowledge networking support of the COST Action CA21106 - COSMIC WISPers in the Dark Universe: Theory, astrophysics and experiments (CosmicWISPers), the COST Action CA22113 - Fundamental challenges in theoretical physics (THEORY-CHALLENGES), the COST Action CA21136 - Addressing observational tensions in cosmology with systematics and fundamental physics (CosmoVerse), the COST Action CA23130 - Bridging high and low energies in search of quantum gravity (BridgeQG), and the COST Action CA23115 - Relativistic Quantum Information (RQI) funded by COST (European Cooperation in Science and Technology). A. \"O. also thanks to EMU, TUBITAK, ULAKBIM (Turkiye) and SCOAP3 (Switzerland) for their support.

\bibliography{ref}

\end{document}